\documentclass[floatfix,aip,cha,10pt,twocolumn,,amsmath,amssymb,superscriptaddress,longbibliography]{revtex4-2}

\usepackage[T1]{fontenc}
\usepackage[utf8]{inputenc}
\usepackage{amsmath,amssymb,amsfonts,amsthm}
\usepackage{graphicx}
\usepackage{units}
\usepackage{dsfont}
\usepackage{xcolor}
\usepackage{algorithmic}
\usepackage{textcomp}
\usepackage{url}

\usepackage[breaklinks]{hyperref}
\usepackage{orcidlink}
\usepackage{todonotes}

\newtheorem{thm}{Theorem}

\newcommand{\matr}[1]{{\boldsymbol{#1}}}
\renewcommand{\vec}[1]{{\boldsymbol{#1}}}
\newcommand{\EE}{\mathcal{E}}
\newcommand{\VV}{\mathcal{V}}

\newcommand{\OO}{\mathcal{F}}

\newcommand{\nn}{{N_\VV}}
\newcommand{\nedge}{{N_\EE}}

\newcommand{\eye}{\mathds{1}}

\begin{document}

\title{Phase Locking in Oscillator Networks with Uncertain and State-Dependent Couplings}

\author{
Iva Ba\v{c}i\'{c}
\email{i.bacic@fz-juelich.de}
\orcidlink{0000-0003-2987-5065}
}
\affiliation{Institute of Climate and Energy Systems: Energy Systems Engineering (ICE-1), Forschungszentrum J\"ulich, 52425 J\"ulich, Germany}
\affiliation{
Institute of Physics Belgrade, University of Belgrade, Belgrade, Serbia}

\author{
Philipp C.~Böttcher\orcidlink{0000-0002-3240-0442}
}
\affiliation{Institute of Climate and Energy Systems: Energy Systems Engineering (ICE-1), Forschungszentrum J\"ulich, 52425 J\"ulich, Germany}

\author{Marc Hunkemöller \orcidlink{0009-0004-6494-1456} }
  \affiliation{Institute of Climate and Energy Systems: Energy Systems Engineering (ICE-1), Forschungszentrum J\"ulich, 52425 J\"ulich, Germany}
  \affiliation{Institute for Theoretical Physics, University of Cologne, 50937 K\"oln, Germany}

\author{
Carsten Hartmann\orcidlink{0009-0007-5067-589X}
}
\affiliation{Institute of Climate and Energy Systems: Energy Systems Engineering (ICE-1), Forschungszentrum J\"ulich, 52425 J\"ulich, Germany}

\author{
David Gross\orcidlink{0000-0001-8888-3028}
}
\affiliation{Institute for Theoretical Physics, University of Cologne,  50937 K\"oln, Germany}

\author{
Andrea Benigni\orcidlink{0000-0002-2475-7003}
}
\affiliation{Institute of Climate and Energy Systems: Energy Systems Engineering (ICE-1), Forschungszentrum J\"ulich, 52425 J\"ulich, Germany}
\affiliation{
RWTH Aachen University, 52062 Aachen, Germany
}

\author{
Dirk Witthaut
\email{d.witthaut@fz-juelich.de}
\orcidlink{0000-0002-3623-5341}
}
\affiliation{Institute of Climate and Energy Systems: Energy Systems Engineering (ICE-1), Forschungszentrum J\"ulich, 52425 J\"ulich, Germany}
\affiliation{Institute for Theoretical Physics, University of Cologne,  50937 K\"oln, Germany}

\begin{abstract}
Phase locking in oscillator networks depends critically on the interaction strengths between coupled units. In many applications, however, these couplings are uncertain, state dependent, or time varying. We derive rigorous conditions for phase locking in Kuramoto-type networks when each coupling strength is known only within prescribed bounds. Building on a convex formulation of the nonlinear fixed-point equations, we obtain edge-wise error bounds for arbitrary feasible approximations that hold uniformly over all admissible couplings. These bounds yield sufficient conditions for the existence of phase-locked states with prescribed phase-difference limits. We illustrate the framework for AC power grids with uncertain voltage magnitudes and adaptive oscillator networks with state-dependent couplings. For the latter, the bounds can be further tightened by exploiting the state dependence of the couplings.
\end{abstract}

\maketitle

\begin{quotation}
The function of many natural and technological systems relies on the synchronization of their constituents. For example, circadian clocks in individual cells must coordinate their rhythms to generate a coherent daily cycle. In electric power grids, generators must rotate with fixed phase differences to sustain the steady transmission of electric power.
Stable synchronization generally requires sufficiently strong coupling between the individual units, making it important to determine whether the available interactions are strong enough to sustain a synchronized state. In many systems, however, the coupling strengths are not precisely known: they may vary in time or depend on other dynamical variables of the system. This raises a basic question: can synchronization still be guaranteed when the coupling strengths are known only approximately? 
We answer this question affirmatively by deriving rigorous conditions for phase locking that hold for all coupling strengths within prescribed bounds.
\end{quotation}

\section{Introduction}

A central mathematical question in the study of synchronization is whether the interactions in a network are sufficiently strong to support a phase-locked state. This question arises in a wide range of systems, including neuronal networks, biological rhythms, coupled lasers, and electric power grids~\cite{pikovsky2003synchronization,arenas2008synchronization,rodrigues2016kuramoto}. In power systems, phase locking of generators is indispensable for stable operation and requires sufficient transmission capacity~\cite{witthaut2022collective}. More generally, sufficiently strong and well-structured coupling is required to sustain phase locking~\cite{dorfler2014survey,cui2026cyclesolvability}, while a loss of effective connectivity can destroy synchrony and, in infrastructure networks, contribute to large-scale failures~\cite{titz2026cascading}.

Most analytical results assume fixed and exactly known coupling strengths. In many real systems, however, the couplings are uncertain, change in time, or co-evolve with the node dynamics, as in temporal and adaptive networks~\cite{holme2012temporal,gross2007adaptive,berner2023adaptive}. Such variations can strongly affect synchronization~\cite{ghosh2022synchronized}. In power grids, for example, the effective coupling strength of a line depends on the voltage magnitudes at its terminal buses, i.e., on the operating state, even for a fixed transmission topology~\cite{witthaut2022collective}. This motivates a formulation in which only the admissible interval of each coupling strength is known.

Here, we address this problem for Kuramoto-type oscillator networks, a paradigmatic framework for synchronization~\cite{kuramoto1975,rodrigues2016kuramoto,dorfler2014survey}. 
Without requiring a detailed model of how the couplings vary, we derive rigorous sufficient conditions for the existence of phase-locked solutions that hold uniformly over the admissible coupling range. 
We illustrate our approach for power grids and adaptive oscillator networks. If the dependence of the couplings on the system state is known, as in adaptive networks, the bounds can be self-consistently tightened.

\section{Fundamentals and Background}

We consider phase-locked states of the Kuramoto model on a complex
network. In a co-rotating frame, these states satisfy
\begin{align}
    \omega_n
    =
    \sum_{m=1}^{\nn}
    K_{nm}\sin(\theta_n-\theta_m),
    \label{eq:kuramoto}
\end{align}
where $\theta_n$ and $\omega_n$ denote the phase and intrinsic frequency
of oscillator $n$, and $K_{nm}=K_{mn}\geq0$ is the coupling strength.
Throughout, we assume that the network is connected and that
$\sum_n\omega_n=0$.

We denote the node and edge sets by $\VV$ and $\EE$, respectively,
with $\nn=|\VV|$ and $\nedge=|\EE|$. After assigning an arbitrary
orientation to every edge, the network topology is encoded in the
node-edge incidence matrix
$\matr E\in\mathbb{R}^{\nn\times\nedge}$. 
The coupling strengths are collected in the diagonal matrix
$\matr K=\operatorname{diag}(K_1,\ldots,K_\nedge)$, the intrinsic frequencies in the vector $\vec \omega$ and the phases in the vector $\vec \theta$.  Equation~\eqref{eq:kuramoto}
then reads
\begin{align}
    \vec\omega
    =
    \matr E\matr K
    \sin(\matr E^\top\vec\theta),
    \label{eq:kuramoto-vector}
\end{align}
where the sine is taken element-wise.

For small phase differences, linearization yields
\begin{align}
    \vec\omega
    =
    \matr E\matr K\matr E^\top\vec\theta^\circ
    =
    \matr L\vec\theta^\circ,
    \label{eq:kuramoto-linear}
\end{align}
where
    $\matr L=\matr E\matr K\matr E^\top$
is the weighted graph Laplacian and ${}^\circ$ denotes an approximate
solution. Since the network is connected,
$\ker(\matr L)=\operatorname{span}\{\vec 1\}$.
Hence, for balanced frequencies $\sum_n\omega_n=0$, the linearized solution is $\vec\theta^\circ=\matr L^+\vec\omega$, where ${}^+$ denotes the Moore--Penrose pseudoinverse.

An equivalent and particularly useful formulation is obtained in terms
of edge flows,
\begin{align}
    \vec f
    =
    \matr K\sin(\matr E^\top\vec\theta),
    \label{eq:def-flow}
\end{align}
such that Eq.~\eqref{eq:kuramoto-vector} reduces to the continuity equation
$
    \vec\omega=\matr E\vec f.
$
This formulation also describes lossless real-power flow in electric grids, where $\omega_n$ is interpreted as the power injection at node $n$~\cite{witthaut2022collective}. Linearization gives
\begin{align}
    \vec f^\circ
    =
    \matr K\matr E^\top\vec\theta^\circ
    =
    \matr K\matr E^\top\matr L^+\vec\omega,
    \label{eq:linear-flow}
\end{align}
corresponding to the DC approximation in power system analysis~\cite{stott2009dc}. Both the exact and linearized flows satisfy the same continuity equation.

On a tree, the continuity equation $\matr E \vec f = \vec \omega$ uniquely determines all edge flows.
In a meshed network, however, the kernel of $\matr E$ has dimension
$\nedge-\nn+1$ and consists of divergence-free cycle flows. We therefore
introduce the projector onto the cycle-flow space~\cite{jafarpour2018synchronization,hartmann2024synchronized},
\begin{align}
    \matr\Pi
    =
    \eye
    -
    \matr K\matr E^\top\matr L^+\matr E .
    \label{eq:def-projector}
\end{align}
For any flow $\vec f$, the component
$
    \vec f_{\rm loop}
    =
    \matr\Pi \vec f
$
is a divergence-free cycle flow,
$
    \matr E\vec f_{\rm loop}
    =
    \vec0.
$
Furthermore,
   $ \matr K\matr\Pi^\top
    =
    \matr K
    -
    \matr K\matr E^\top
    \matr L^+\matr E\matr K
        =
    \matr\Pi\matr K$,
and $\matr\Pi$ is an orthogonal projector with respect to the inner product
 $   \langle\vec\xi,\vec\zeta\rangle_K
    :=
    \vec\xi^\top\matr K^{-1}\vec\zeta $.

On meshed networks, phase-locked solutions need not be unique. In particular, the phase differences around a cycle may accumulate to an integer multiple of $2\pi$, giving rise to distinct winding sectors or \emph{twisted states}~\cite{manik2017cycle,delabays2016multistability,jafarpour2022flow}. These states are characterized by an integer winding vector $\vec z\in\mathbb{Z}^{\nedge-\nn+1}$ associated with a basis of independent cycles. In the following, we restrict attention to the zero-winding sector $\vec z=\vec 0$, which contains the phase-locked state continuously connected
to the small-angle solution.

\section{Rigorous results}

We now derive our main results on the existence of phase-locked states in networks with uncertain coupling strengths. More precisely, we assume that for every edge $e$ only lower and upper bounds
\begin{align}
    0<\underline K_e\le K_e\le\overline K_e,
    \label{eq:K-bounds}
\end{align}
are known. The central question is whether rigorous conclusions can still be drawn from the DC approximation, or from any other feasible approximation, without knowing the exact values of the couplings.

Our analysis builds on an a posteriori error bound derived in Ref.~\cite{hartmann2024synchronized,bacic2026linear} for fixed, known coupling strengths $\matr K$.

\begin{thm}
\label{thm:known-K}
Let $\vec\theta^*$ be a solution of the nonlinear fixed-point equation~\eqref{eq:kuramoto-vector} with zero winding and
$
    |\theta_n^*-\theta_m^*|\le \pi/2
$
for all $(n,m)\in\EE$. Let
$
    \vec f^*
    =
    \matr K\sin(\matr E^\top\vec\theta^*)
$
be the associated flow vector, and let $\vec f^\circ$ be any approximation satisfying the continuity equation
$
    \matr E\vec f^\circ=\vec\omega .
$
Then the approximation error
$
    \vec\xi=\vec f^\circ-\vec f^*
$
satisfies
\begin{align}
    \vec\xi^\top\matr K^{-1}\vec\xi
    \le
    \vec g(\vec f^\circ)^\top\vec\xi,
    \label{eq:inequality-from-Taylor}
\end{align}
where the components of $\vec g$ are 
    $g_e(\vec f)
    = \arcsin(f_e/K_e).$
\end{thm}

The key idea behind Theorem~\ref{thm:known-K} is to reformulate the nonlinear fixed-point equation~\eqref{eq:kuramoto-vector} as a convex optimization problem in edge space, subject to the constraint $\matr E\vec f=\vec\omega$~\cite{hartmann2024synchronized,chertkov2026gigawatts}. The function $\vec g(\vec f)$ is the gradient of the objective function
\begin{align}
    \label{eq:objective-fun}
    \OO(\vec f) &= \sum_{e \in \EE} \OO_e(f_e), \\
     \OO_e(f_e) &= 
     f_e \arcsin \left( \frac{f_e}{K_e} \right)
     + \sqrt{K_e^2 - f_e^2},\label{opt:objfunc}
\end{align}
while $\matr K^{-1}$ provides a lower bound on its Hessian. Equation~\eqref{eq:inequality-from-Taylor} then follows from Taylor's theorem~\cite{hartmann2024synchronized}. 

The convex problem has a unique minimizer $\vec f^*$ for every fixed coupling matrix $\matr K$. If $|f_e^*|<K_e$ on all edges, this minimizer corresponds to a zero-winding solution $\vec\theta^*$ of the nonlinear fixed-point equations, obtained by inverting Eq.~\eqref{eq:def-flow}. The result can be generalized to nonzero winding sectors~\cite{hartmann2024synchronized}, corresponding to twisted states.

Theorem~\ref{thm:known-K} provides a global a posteriori error bound for fixed and perfectly known coupling strengths. In the present setting, two extensions are required. First, both sides of Eq.~\eqref{eq:inequality-from-Taylor} depend on the unknown coupling matrix $\matr K$, and we therefore seek bounds that hold uniformly for all $\matr K$ satisfying Eq.~\eqref{eq:K-bounds}. Second, phase-locking constraints are imposed edge by edge, so a single global bound is not sufficient. We therefore derive individual error bounds for every edge, which will allow us to formulate robust conditions for phase locking under uncertain coupling strengths.

We begin by bounding the quantities entering Eq.~\eqref{eq:inequality-from-Taylor} using only the admissible coupling intervals. Let $\vec f^*$ denote the exact nonlinear flow and $\vec f^\circ$ an arbitrary approximation satisfying $\matr E\vec f^\circ=\vec\omega$. Without loss of generality, we choose the orientation of every edge such that $f_e^\circ\ge0$. 
Assuming $f_e^\circ<\underline K_e$, monotonicity of the arcsine implies
\begin{align}
    0
    \le
    \underbrace{
    \arcsin\!\left(\frac{f_e^\circ}{\overline K_e}\right)
    }_{=\underline g_e}
    \le
    g_e(\vec f^\circ)
    \le
    \underbrace{
    \arcsin\!\left(\frac{f_e^\circ}{\underline K_e}\right)
    }_{=\overline g_e}.
    \label{eq:gradient-bounds}
\end{align}
Thus, the unknown gradient is bounded component-wise by the vectors $\underline{\vec g}$ and $\overline{\vec g}$.
We further collect the coupling bounds in the diagonal matrices
\begin{align}
    \underline{\matr K}
    &=
    \operatorname{diag}(\underline K_1,\ldots,\underline K_{\nedge}),
    &
    \overline{\matr K}
    &=
    \operatorname{diag}(\overline K_1,\ldots,\overline K_{\nedge}),
    \nonumber
\end{align}
and define the Laplacian and cycle-flow projector associated with the upper coupling bounds,
\begin{align}
    \overline{\matr L}
    &=
    \matr E\overline{\matr K}\matr E^\top,
    \qquad
    \overline{\matr\Pi}
    =
    \eye
    -
    \overline{\matr K}\matr E^\top
    \overline{\matr L}^{+}\matr E .
    \label{eq:uncertain-projector}
\end{align}
For later use, we define
$
    {\matr A}
    :=
    \overline{\matr\Pi}\,\overline{\matr K}.
$
Using
$\overline{\matr K}\,\overline{\matr\Pi}^{\top}
=
\overline{\matr\Pi}\,\overline{\matr K}$,
the matrix ${\matr A}$ is symmetric.
We further introduce its element-wise positive and negative parts,
$A^+_{e,e'}=\max(\overline A_{e,e'},0)$ and 
$A^-_{e,e'}=\max(-\overline A_{e,e'},0)$
such that
\begin{align}
    {\matr A}
    =
    {\matr A}^{+}
    -
    {\matr A}^{-}.
    \label{A-plus-minus}
\end{align}
We then get the following result.

\begin{thm}
\label{thm:errorbound-upperlower}
Let $\vec\theta^*$ be a solution of the nonlinear fixed-point equation~\eqref{eq:kuramoto-vector} with zero winding and
$|\theta_n^*-\theta_m^*|\le \pi/2$ for all $(n,m)\in\EE$ and $\vec f^* = \matr K\sin(\matr E^\top\vec\theta^*)$.
For any approximation $\vec f^\circ$ satisfying $\matr E \vec f^\circ = \vec \omega$ and $|f^\circ_e| < \underline K_e$ for all $e \in \EE$, the error  $\vec \xi = \vec f^\circ -\vec f^*$ is bounded as
\begin{align}
      \underline \kappa_e - \overline \chi_e \le   
      \xi_e
      \le \overline \kappa_e + \overline \chi_e ,
\end{align}
where
\begin{align}
    \overline \kappa_e &= \frac{1}{2} \left( \vec w_e^\top   {\matr A}^{+}\overline{\vec g}
    -
    \vec w_e^\top {\matr A}^{-}\underline{\vec g}
    \right),
    \\
    \underline \kappa_e &= \frac{1}{2} \left( \vec w_e^\top   {\matr A}^{+}\underline{\vec g}
    -
    \vec w_e^\top  {\matr A}^{-} \overline{\vec g} \right),
    \\
    \overline{\chi}_e^2 &= \frac{\overline{K}_e \overline{\Pi}_{e,e}}{4} \left(
    \overline{\vec g}^{\,\top}
    {\matr A}^{+}
    \overline{\vec g}
    -
    \underline{\vec g}^{\,\top}
    {\matr A}^{-}
    \underline{\vec g} \right),
\end{align}
for every edge $e$ that is not a bridge. Here, $\vec w_e$ denotes the $e$-th standard basis vector,
$(\vec w_e)_{e'}=\delta_{e,e'}$. For every bridge, $\underline \kappa_e = \overline \kappa_e = \overline \chi_e = 0$.
\end{thm}

\begin{proof}
If $e$ is a bridge, every cycle flow vanishes on $e$. Since $\vec\xi\in\ker(\matr E)$, it follows that $\xi_e=0$, and hence $\underline\kappa_e=\overline\kappa_e=\overline\chi_e=0$.

For all other edges, we start from the bound
\eqref{eq:inequality-from-Taylor}. Since
$\matr K\preceq\overline{\matr K}$, we have
$\matr K^{-1}\succeq\overline{\matr K}^{-1}$ and therefore
\begin{align}
    \vec\xi^\top\overline{\matr K}^{-1}\vec\xi
    -
    \vec g^\top\vec\xi
    \le0 .
    \label{eq:uncertain-global-bound}
\end{align}

We now determine which values of $\xi_e$ are compatible with this bound. For every edge, consider
\begin{align}
    R_e(\xi_e)
    ={}&
    \min_{\vec\zeta}
    \left[
        \vec\zeta^\top\overline{\matr K}^{-1}\vec\zeta
        -
        \vec g^\top\vec\zeta
    \right]
    \nonumber\\
    &\mathrm{s.t.}\qquad
    \matr E\vec\zeta=\vec0,
    \qquad
    \zeta_e=\xi_e .
    \label{eq:optimization-R}
\end{align}
A value $\xi_e$ is compatible with
Eq.~\eqref{eq:uncertain-global-bound} only if
$R_e(\xi_e)\le0$.

The quadratic optimization problem~\eqref{eq:optimization-R} can be
solved in closed form. The minimizer is
$    \vec\zeta^*
    =
    \frac{1}{2}
    \overline{\matr\Pi}\,
    \overline{\matr K}
    \left(
        \vec g-\mu\vec w_e
    \right),$
with the Lagrange multiplier
$
    \mu
    =
    (
        \vec w_e^\top
        \overline{\matr\Pi}\,
        \overline{\matr K}\vec g
        -
        2\xi_e
    )/(
        \overline K_e\overline\Pi_{e,e}
    ) .
$
Using the properties of the projector, the condition
$R_e(\xi_e)\le0$ can then be recast as
\begin{align}
   \left(
      \vec w_e^\top
      \overline{\matr\Pi}\,
      \overline{\matr K}\vec g
      - 2 \xi_e
   \right)^2
   \le
   \overline K_e\overline\Pi_{e,e}
   \,
   \vec g^\top
   \overline{\matr\Pi}\,
   \overline{\matr K}\vec g ,
   \label{eq:inequality-mu-eta-bounds2}
\end{align}
which implies 
\begin{align}
      \xi_e &\le 
      \frac{1}{2}
      \vec w_e^\top
      \overline{\matr\Pi}\,
      \overline{\matr K}\vec g
      + \left(
   \frac{\overline K_e\overline\Pi_{e,e}}{4}
   \,
   \vec g^\top
   \overline{\matr\Pi}\,
   \overline{\matr K}\vec g \right)^{1/2},
   \nonumber \\
    \xi_e &\ge 
      \frac{1}{2}
      \vec w_e^\top
      \overline{\matr\Pi}\,
      \overline{\matr K}\vec g
      - \left(
   \frac{\overline K_e\overline\Pi_{e,e}}{4}
   \,
   \vec g^\top
   \overline{\matr\Pi}\,
   \overline{\matr K}\vec g \right)^{1/2}.
   \label{eq:inequality-mu-eta-bounds3}
\end{align}

It remains to eliminate the unknown gradient $\vec g$.
Using the component-wise bounds
$\underline{\vec g}\le\vec g\le\overline{\vec g}$
and the decomposition~\eqref{A-plus-minus}, we obtain
\begin{align*}
    \vec w_e^\top \!\! \left( {\matr A}^{+}\underline{\vec g}
    -
    {\matr A}^{-} \overline{\vec g} \right)
    &\le
    \vec w_e^\top
    \overline{\matr\Pi}\,
    \overline{\matr K}\vec g
    \le
    \vec w_e^\top \!\! \left( {\matr A}^{+}\overline{\vec g}
    -
    {\matr A}^{-}\underline{\vec g}
    \right)
\end{align*}
and
 $   \vec g^\top
    \overline{\matr\Pi}\,
    \overline{\matr K}\vec g
    \le
    \overline{\vec g}^{\,\top}
    {\matr A}^{+}
    \overline{\vec g}
    -
    \underline{\vec g}^{\,\top}
    {\matr A}^{-}
    \underline{\vec g}.$ 
Substitution into Eq.~\eqref{eq:inequality-mu-eta-bounds3}
yields the claimed bounds.
\end{proof}

We now extend the a posteriori error bounds to an existence result for the nonlinear fixed-point equation~\eqref{eq:kuramoto-vector}. We recall that Theorem~\ref{thm:known-K} is based on a convex optimization problem in edge space whose unique minimizer $\vec f^*$ exists for every fixed coupling matrix $\matr K$. A phase-locked state $\vec\theta^*$ solving Eq.~\eqref{eq:kuramoto-vector} is obtained whenever the minimizer lies in the principal domain, $|f_e^*|<K_e$ for all $e\in\EE$. Importantly, the error bounds derived above apply to the minimizer $\vec f^*$ irrespective of whether it corresponds to a phase-locked state.  
Therefore, starting from any feasible approximation $\vec f^\circ$, Theorem~\ref{thm:errorbound-upperlower} provides a rigorous region of trust containing $\vec f^*$. If this entire region lies inside the principal domain on every edge, then $\vec f^*$ necessarily corresponds to a phase-locked solution of Eq.~\eqref{eq:kuramoto-vector}.

\begin{thm}[Robust existence of phase-locked states]
\label{thm:robust-existence}
Let $\vec f^\circ$ be a feasible approximation satisfying $\matr E\vec f^\circ=\vec\omega$, and let 
$0<\gamma_e<\pi/2$ be a prescribed limit on the phase difference for each edge $e\in\EE$.
If, for every edge $e\in\EE$,
\begin{align}
    f_e^\circ-\overline\kappa_e-\overline\chi_e
    &\ge
    -\underline K_e\sin(\gamma_e),
    \label{eq:robust-existence-lower}
    \\
    f_e^\circ-\underline\kappa_e+\overline\chi_e
    &\le
    \underline K_e\sin(\gamma_e),
    \label{eq:robust-existence-upper}
\end{align}
then, for every realization of the coupling strengths satisfying
$ \underline K_e\le K_e\le\overline K_e$,
the nonlinear fixed-point equation~\eqref{eq:kuramoto-vector} admits a zero-winding solution $\vec\theta^*$ satisfying
\begin{align}
    |\theta_n^*-\theta_m^*|
    \le\gamma_e
    \qquad
    \forall e=(n,m)\in\EE .
\end{align}
\end{thm}

\begin{proof}
Theorem~\ref{thm:errorbound-upperlower} implies
\begin{align}
    f_e^\circ-\overline\kappa_e-\overline\chi_e
    \le
    f_e^*
    \le
    f_e^\circ-\underline\kappa_e+\overline\chi_e .
\end{align}
Conditions~\eqref{eq:robust-existence-lower} and
\eqref{eq:robust-existence-upper} therefore yield
\begin{align}
    |f_e^*|
    \le
    \underline K_e\sin(\gamma_e)
    \le
    K_e\sin(\gamma_e)
    <
    K_e,
\end{align}
for every admissible realization of $\matr K$.
Hence the minimizer $\vec f^*$ lies in the principal domain and
corresponds to a zero-winding solution of
Eq.~\eqref{eq:kuramoto-vector}. Since
$f_e^*=K_e\sin(\theta_n^*-\theta_m^*)$ and
$\gamma_e<\pi/2$, it follows that
$|\theta_n^*-\theta_m^*|\le\gamma_e$.
\end{proof}

\section{Examples and applications}

\subsection{Power grids with uncertain voltage magnitudes}

We first illustrate the robust existence criterion for AC power grids. Neglecting Ohmic losses, the nodal real-power balance is
\begin{align}
    p_n
    =
    \sum_m b_{nm}v_nv_m\sin(\theta_n-\theta_m),
    \label{eq:realpower-balance}
\end{align}
where $v_n$ and $\theta_n$ denote the voltage magnitude and phase angle at node $n$, and $b_{nm}$ is the line susceptance. 
Even for a fixed transmission topology, the effective coupling strength $K_{nm}=b_{nm}v_nv_m$ depends on the operating state through the voltage magnitudes.

\begin{figure}
\includegraphics[width=0.95\columnwidth]{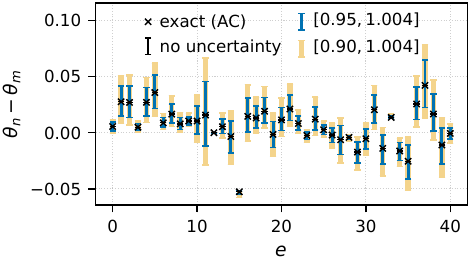}
\caption{Error bounds under uncertain voltage magnitudes for the \texttt{case30} test grid. Error bars show the bounds on the phase differences $\theta_n-\theta_m$ across each line $e=(n,m)$, derived from Theorem~\ref{thm:errorbound-upperlower} for exactly known voltage magnitudes (black) and for load-bus voltages known only to satisfy $v_n\in[\underline v_n,\overline v_n]$, with the intervals indicated in the legend (blue,  light orange). Crosses mark the numerically exact solution. Generator buses are controlled to have fixed voltages.}
\label{fig:uncertainvoltages}
\end{figure}

In normal operation, voltage magnitudes remain within prescribed bounds,
   $ \underline v_n
    \le
    v_n
    \le
    \overline v_n$,
typically close to the nominal value $v_n^{\rm ref}=1$ in per-unit
coordinates. At generator buses, the voltage is fixed, $\underline v_n=\overline v_n=v_n^{\rm ref}$. At load buses, we obtain an upper bound from the reactive power balance: if all load buses draw reactive power, the load-bus voltages cannot exceed the open-circuit voltages $\hat{\vec v}_\mathcal{L}=-\matr{B}_{\mathcal{L}\mathcal{L}}^{-1}\matr{B}_{\mathcal{L}\mathcal{G}}\vec v_\mathcal{G}$, provided that $-\matr{B}_{\mathcal{L}\mathcal{L}}$ is positive definite~\cite{simpson2016voltage}. Here, $\matr{B}$ is the nodal susceptance matrix including shunts, partitioned into load ($\mathcal{L}$) and generator ($\mathcal{G}$) buses. The lower bounds $\underline v_n$ are assumed according to operational voltage limits. These bounds directly imply
\begin{align}
    \underline K_{nm}
    &=
    b_{nm}\underline v_n\underline v_m,
    &
    \overline K_{nm}
    &=
    b_{nm}\overline v_n\overline v_m.
\end{align}
Operational security further requires bounded phase-angle differences. Theorem~\ref{thm:robust-existence} can therefore be used to guarantee the existence of a phase-locked state satisfying
 $   |\theta_n-\theta_m|
    \le
    \gamma_{nm}$
for all lines, without requiring the exact voltage magnitudes.

Figure~\ref{fig:uncertainvoltages} demonstrates the  approach for a modified \textsc{Matpower} \texttt{case30} test system~\cite{matpower2011}. Ohmic losses are neglected by setting all branch resistances and shunt conductances to zero. Starting from the DC approximation $\vec f^\circ$ with nominal load-bus voltages $v_n=1$, we apply Theorem~\ref{thm:errorbound-upperlower} using only the prescribed voltage intervals. The resulting bounds on the flows are converted into bounds on the phase differences using the worst-case couplings $\underline K_e$, $\overline K_e$ and the monotonicity of the arcsine. These bounds are compared with the numerically exact solution of Eq.~\eqref{eq:realpower-balance}.
The bounds remain informative despite the uncertainty in the voltage magnitudes and provide an a priori guarantee for phase-locked operation.

\subsection{State-dependent couplings in adaptive oscillator networks}

As a second example, we consider adaptive oscillator networks in which the coupling strengths evolve in response to the phases. A standard model is given by Kuramoto oscillators with Hebbian adaptation~\cite{seliger2002plasticity,aoki2009co, bronski2017stability, bacic2018noise, juttner2023complex, cestnik2025continuum},
\begin{align}
    \dot\theta_n
    &=
    \omega_n
    +
    \sum_m K_{nm}\sin(\theta_m-\theta_n), \label{eq:adaptive-theta}
    \\
    \dot K_{nm}
    &=
    \varepsilon
    \left[
        K_0+\alpha\cos(\theta_n-\theta_m)-K_{nm}
    \right],
    \label{eq:adaptive-K}
\end{align}
where $\varepsilon\ll1$ describes the adaptation rate. Couplings between nearly synchronized oscillators are strengthened, whereas larger phase differences weaken the interaction. We investigate whether the adaptive network has a stable phase-locked state, where $\dot \theta_n = 0$ and $\dot K_{nm} = 0$, and the couplings are determined by the phases,
\begin{align}
    K_{nm} = K_0+ \alpha \cos(\theta_n-\theta_m). \label{eq:adapted-coupling}
\end{align}

\textbf{A priori bounds.} For $K_0>\alpha>0$, the interval
\begin{align}
    K_0-\alpha
    \le K_{nm}(t)
    \le K_0+\alpha
    \label{eq:adaptive-bounds}
\end{align}
is positively invariant, provided the initial coupling strengths lie inside this interval. Hence, the adaptive dynamics directly provide the coupling bounds required by Theorem~\ref{thm:robust-existence}, without any knowledge of the phases. However, these bounds may be weak because they must be applicable to a network in which all couplings have the smallest possible value $K_0-\alpha$.

\textbf{Self-consistent bounds.} We derive a stronger result based on the state dependence of the couplings, Eq.~\eqref{eq:adapted-coupling}. If $|\theta_n-\theta_m|\le\gamma_e$ in the phase-locked state, then
\begin{align}
\underline{K}_e= K_0 + \alpha\cos \gamma_e\le K_e\le K_0 + \alpha =  \overline{K}_e.\label{eq:self-consistent-adaptive}
\end{align}
If Theorem~\ref{thm:robust-existence} with these bounds certifies angles of at most $\gamma_e$, the assumption is self-consistent, because the couplings generated by any certified phase-locked state also satisfy~\eqref{eq:self-consistent-adaptive}. Brouwer's fixed-point theorem guarantees a fixed point of the adaptive network with $|\theta_n-\theta_m|\le\gamma_e$. We obtain such angles by iterating this step, starting
from $\gamma_e=0$. By extending the analysis of Ref.~\cite{bronski2017stability} to $K_0>0$, this equilibrium is linearly stable if 
\begin{align}
\cos \gamma_e > \frac{-K_0+\sqrt{K_0^2+8\alpha^2}}{4\alpha} \qquad \forall e \in \EE.\label{eq:stability}
\end{align}

\begin{figure}
\includegraphics[width=0.95\columnwidth]{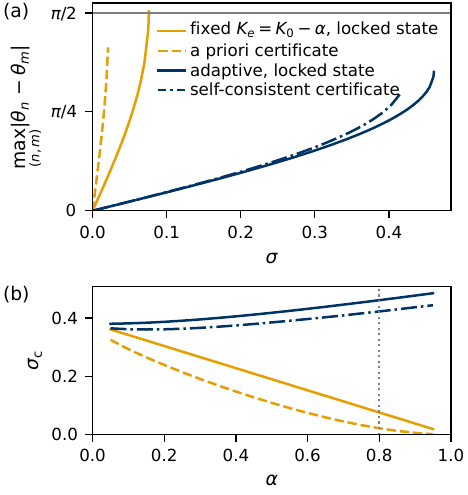}
\caption{Certified phase locking in an adaptive ring of $N=50$ oscillators with $K_0 = 1$.
(a) Largest phase difference versus the frequency spread $\sigma$ for $\alpha=0.8$. Each certificate bounds the corresponding solid line from above, up to $\sigma_c$ where it fails.
(b) Critical frequency spread $\sigma_\mathrm{c}$ versus $\alpha$, beyond which the locked state ceases to exist (solid) or is no longer certified (dashed: a priori, dash-dotted: self-consistent bounds). Below the dash-dotted line, the certified equilibrium is also linearly stable by Eq.~\eqref{eq:stability}. The dotted line marks $\alpha=0.8$.}
\label{fig:adaptive}
\end{figure}

We compare these two approaches on a ring of $N=50$ oscillators with $K_0=1$ (Fig.~\ref{fig:adaptive}). The intrinsic frequencies are chosen from a single realization of i.i.d.\
standard normal variables, shifted to zero mean and multiplied by $\sigma$. We compute the phase-locked states using Newton's method, continuing the solution in $\sigma$ from $\sigma=0$ and accepting only stable states. 
With a priori bounds, the certificate fails at small $\sigma$, below the threshold of the non-adaptive network, cf. Fig.~\ref{fig:adaptive}(a), and deteriorates further with increasing $\alpha$ as the lowest admissible coupling $K_0-\alpha$ decreases, cf. Fig.~\ref{fig:adaptive}(b). With self-consistent bounds, the certificate closely follows the adaptive equilibrium, certifying stable locking almost up to its fold, and, beyond small $\alpha$, it improves with $\alpha$. For sufficiently strong adaptation, it certifies phase locking where the network with fixed couplings $K_0$ does not have a phase-locked solution, thus capturing how adaptation enhances synchronization.

\section{Conclusion}

We derived rigorous conditions for phase locking in the Kuramoto model on networks whose coupling strengths are known only within prescribed bounds. For power grids with uncertain voltage magnitudes, these guarantees remain informative under realistic voltage intervals. For adaptive networks, the bounds can be made even tighter by exploiting the state dependence of the couplings through a self-consistent choice of bounds.

An important next step is to extend the framework to the Sakaguchi-Kuramoto model, which arises both in power grids with Ohmic losses, and in modeling adaptive networks. Furthermore, since the tightness of the bounds depends on the network topology through the cycle-flow projector, it would be valuable to obtain a systematic understanding of how network structure influences the robustness of phase locking under coupling uncertainty.

\section*{Acknowledgments}

The work of I.~Ba\v{c}i\'{c} and D.~Witthaut was supported by the German Federal Ministry of Research, Technology and Space (Bundesministerium für Forschung, Technologie und Raumfahrt) via the grant number 03SF0751. The authors acknowledge the use of Claude (Anthropic) for assistance with code development.

\section*{AUTHOR DECLARATIONS}

\subsection*{Conflict of Interest}

The authors have no conflicts to disclose.

\subsection*{Author Contributions}

\textbf{Iva Ba\v{c}i\'{c}}: Conceptualization (equal); Formal analysis (equal); Investigation (equal); Methodology (equal); Software (equal); Validation (equal); Visualization (equal); Writing -- original draft (equal); Writing -- review \& editing (equal). \textbf{Philipp C. Böttcher}: Investigation (supporting); Software (equal); Validation (equal); Visualization (equal); Writing -- review \& editing (supporting). \textbf{Marc Hunkemöller}: Investigation (supporting); Software (supporting); Validation (supporting); Visualization (supporting); Writing -- review \& editing (supporting).
\textbf{Carsten Hartmann}: Conceptualization (supporting); Formal analysis (supporting); Methodology (supporting); Validation (supporting); Writing -- review \& editing (supporting). \textbf{David Gross}: Conceptualization (supporting); Methodology (supporting); Writing -- review \& editing (supporting). \textbf{Andrea Benigni}: Funding acquisition (supporting); Supervision (supporting); Writing -- review \& editing (supporting). \textbf{Dirk Witthaut}: Conceptualization (equal); Formal analysis (equal); Methodology (equal); Supervision (lead); Project administration (lead); Writing -- original draft (lead); Writing -- review \& editing (equal).

\subsection*{Data  Availability}

Supporting code is available at: \url{https://jugit.fz-juelich.de/p.boettcher/uncertain-couplings/}.

%

\end{document}